\documentclass[twocolumn]{article}

\usepackage[letterpaper, margin=0.75in]{geometry}
\usepackage{graphicx}
\usepackage{amsmath}
\usepackage{amsfonts}
\usepackage{authblk} 
\usepackage{cite}     
\usepackage{array}    
\usepackage{tabularx}
\usepackage{makecell}
\usepackage{dblfloatfix}
\usepackage{microtype}

\usepackage{xcolor} 

\newcommand{\corr}{}

\title{\bfseries{Intelligent Mode Sorting in Turbulence with Task-Dependent Optical Neural Networks}}

\author[]{Christopher R. Rawlings} 
\author[\corr]{Mitchell A. Cox}

\affil[]{School of Electrical and Information Engineering, University of the Witwatersrand, Johannesburg, South Africa}
\affil[\corr]{e-mail: mitchell.cox@wits.ac.za}

\date{} 

\begin{document}

\maketitle

\begin{abstract}
The practical deployment of high-capacity free-space optical communication is fundamentally limited by atmospheric turbulence, a challenge that conventional mode-sorting techniques have failed to overcome. While engineered optical computers like diffractive networks offer a potential solution, their design complexity remains a significant barrier. Here, we introduce and experimentally validate a methodology for task-dependent hardware design, demonstrating a simple, continuous-wave-driven multimode fibre reservoir that is physically configured to solve this specific, high-impact problem. We first establish a set of physical design principles, showing that recurrent dynamics are optimal for structural data (MNIST), whereas high mode-mixing is superior for textural data (FMNIST). By treating turbulence-induced wavefront distortion as a complex textural feature, we configure a physically optimised reservoir for classifying orbital angular momentum (OAM) modes. In moderate to high-turbulence regimes, our system outperforms an ideal modal decomposition by an average of  20.32~$\pm$~3.00\%, succeeding precisely where the conventional approach fails. This work not only demonstrates a practical, low-complexity solution for turbulence mitigation but also reframes the optical receiver as a physical likelihood processor, thereby offering a path towards significantly reduced digital signal processing burdens by offloading much of the computational load to the physical optical front-end. We establish a framework for co-designing physical hardware with computation, enabling simpler, more robust optical machine learning systems tailored for real-world challenges.
\end{abstract}

\section{Introduction}

The pursuit of faster, more energy-efficient computation has reinvigorated the field of optical computing \cite{vanderSande2017photonic, momeni2024training}. A key frontier is the development of systems that can process optical data natively, circumventing electronic bottlenecks in applications such as free-space optical (FSO) communications. High-capacity FSO links, which often employ spatial modes of light like orbital angular momentum (OAM) for multiplexing, have demonstrated extraordinary channel capacities of up to 1~Pb/s in laboratory settings \cite{wang2014fso1pb}, and 400~Gb/s over only 120~m free-space \cite{Ren2016}. This transition to real-world environments is obstructed by a fundamental barrier: atmospheric turbulence, which severely distorts the wavefront and corrupts the encoded information \cite{Trichili2020Roadmap, cox2020slt}. Mitigation with optimised optical modes, adaptive optics, or advanced digital signal processing is complex and costly. Recent attempts to employ digital machine learning, such as convolutional neural networks, for classifying distorted modes introduce a significant power and latency burden. This burden arises from the analogue-to-digital and digital-signal-processing pipeline, ultimately rendering these approaches impractical for real-time communication \cite{Wang2022_OAM_FSO_Review,Amirabadi2024_Survey_ML_DL_OpticalComms}. This creates a distinct opportunity for a new class of optical receiver front-end: a passive, low-latency "intelligent mode sorter" capable of robust classification at the speed of light.

Addressing this challenge has inspired two distinct design philosophies for optical computers. The first, a paradigm of "prescriptive design," involves meticulously engineering the system's degrees of freedom to impose a specific, pre-calculated mathematical function onto the light field. This approach, exemplified by diffractive deep neural networks \cite{lin2018all, Mengu2020} and complex photonic integrated circuits \cite{wu2021programmable, ashtiani2022chip, xia2025chip}, offers remarkable precision and a direct path to implementing known algorithms. An alternative philosophy, which we explore here, is one of "emergent design." This paradigm seeks to harness the immense and intrinsic computational power of a complex physical system \cite{cucchi2022hands}, guiding its emergent properties through the tuning of a few macroscopic parameters. For problems involving stochastic, high-dimensional inputs like atmospheric turbulence, this approach offers a compelling path towards solutions that are simpler, more efficient, and more elegant.

The principle of harnessing intrinsic complexity is generalisable. It relies on implementing a physical "random projection," a process achievable in various complex optical systems, from multimode fibres to other scattering media, whose stability can be engineered for specific applications. While the physical medium itself may have environmental sensitivities, a trained machine learning system can learn to be robust to the specific, task-relevant distortions. By formalising this process within the Extreme Learning Machine (ELM) framework \cite{huang2006extreme}, we can understand the complex medium as a powerful, high-dimensional, fixed random projector. This theoretical grounding reveals a crucial practical advantage: the system can be trained with a computationally trivial, one-shot analytical step, bypassing the need for iterative backpropagation. It also allows for high performance with a simple, low-power continuous-wave laser, where the square-law detection at the camera provides a surprisingly sufficient nonlinearity for these complex tasks. By formalising this physical system within an established machine learning framework, we aim to create a bridge between the photonics and computer science communities, inviting theoretical insights to inform physical hardware design and vice-versa.

However, the full potential of physical reservoirs has been limited by their treatment as static, opaque "black boxes." Our work aims to light up the inside of this box. We propose and experimentally demonstrate a methodology of task-dependent hardware co-design, arguing that the optimal physical configuration of the reservoir is not universal but is dictated by the fundamental structure of the data it must process. To prove this principle, we first construct a design map by systematically characterising how the reservoir's physical parameters—such as mode mixing and recurrent feedback—affect its ability to classify data defined by simple structures (MNIST) versus complex textures (FMNIST).

With these design principles established, we return to the grand challenge of MDM FSO communications. Treating turbulence-induced wavefront distortion as a complex textural feature, we apply our methodology to configure a physically optimised reservoir for classifying OAM modes. The system demonstrates remarkable resilience, outperforming a conventional, power-normalised modal decomposition benchmark by an average of  20.32~$\pm$~3.00~\% in moderate to high-turbulence regimes where robust classification is most critical. This result is more than a better classifier; it is a proof-of-concept for a new optical receiver architecture. We show that the reservoir functions as a physical likelihood processor, whose output provides the rich, soft information required for state-of-the-art soft-decision forward error correction (FEC). Our work thus establishes a framework for co-designing physical hardware with computation, demonstrating a path towards simpler, more robust optical systems where complex digital processing is offloaded to passive, speed-of-light computation.

\section{Results}
\label{sec:results}
\paragraph{Experiment Design}

\begin{figure}[t]
    \centering
    \includegraphics[width=\columnwidth]{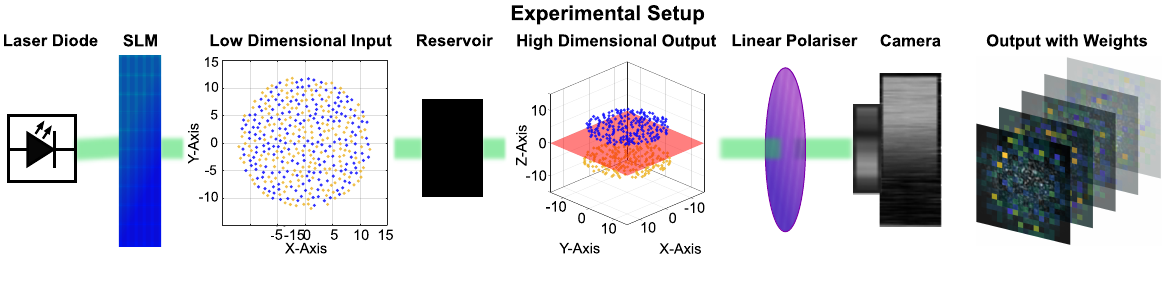} 
    \caption{\textbf{Experimental Setup.} An SLM encodes the input (MNIST, FMNIST and OAM modes in turbulence) onto a continuous-wave laser. The reservoir projects these inputs to a higher dimensional space to form speckle patterns. The linear polariser is added for complexity. The square law detection and saturation of the camera adds a quadratic nonlinearity to the system. The captured speckle patterns are used to train the output weights used for classification. }
    \label{fig:ExpSetup}
\end{figure}

To solve a complex, real-world classification problem, we must first understand how to configure the physical hardware for optimal performance. The experimental system is shown in Fig.~\ref{fig:ExpSetup}. A continuous-wave laser illuminates a spatial light modulator (SLM), which encodes the input data onto the phase and/or amplitude of the beam. The first-order diffracted beam is isolated using a spatial filter and injected into the physical reservoir. At the output of the reservoir, a linear polariser is placed before the camera. The polariser acts as a spatial filter, converting the complex, spatially varying state of polarisation of the speckle field into additional intensity variations, thereby enriching the structure of the captured pattern.

We hypothesise that the ideal reservoir configuration is strongly dependent on the physical nature of the data being classified. To build a set of design principles, we systematically characterised the performance of our fibre reservoir on two canonical datasets that encode information differently: MNIST, which consists of digits defined by simple, structural edges, and FMNIST, which consists of clothing items defined by complex, textural features.

\paragraph{Mapping Reservoir Physics to Data Structure}
\begin{figure*}[ht]
    \centering
    \includegraphics[width=\textwidth]{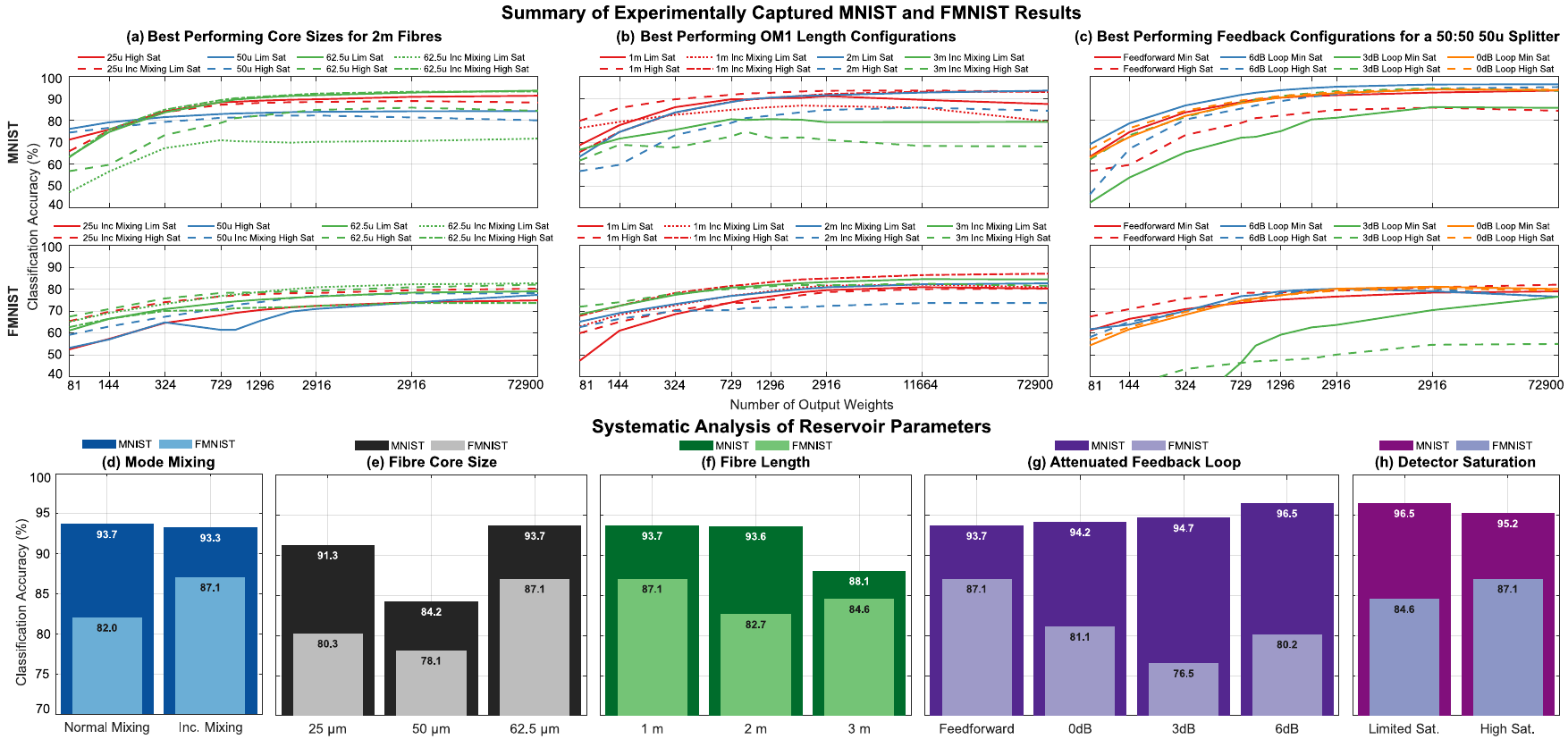} 
    \caption{\textbf{Establishing the principles of task-dependent reservoir design.} To build a design map, the peak classification accuracy for data defined by structure (MNIST) and texture (FMNIST) is systematically tested against several physical reservoir parameters in (a) to (c). The impacts of these parameters are summarised in (d) induced mode mixing, (e) fibre core size, (f) fibre length, (g) a feedback loop (with 0~dB, 3~dB and 6~dB of attenuation), and (h) detector saturation. The results reveal clear design trade-offs, for instance, feedback is highly beneficial for structural data but detrimental to textural data. This demonstrates that the optimal physical hardware configuration is critically dependent on the input data, forming the basis of our design methodology. }
    \label{fig:ParameterAnalysis}
\end{figure*}

Our investigation reveals a clear, task-dependent relationship between the reservoir's physical parameters and classification accuracy. 

First, we increased the spatial complexity of the reservoir through externally induced mode mixing, achieved by applying weighted pellets to the fibre coil to increase micro-bending. These are best illustrated by the 1~m OM1 comparison and 62.5~$\mu$m comparisons in Fig.~\ref{fig:ParameterAnalysis}a and Fig.~\ref{fig:ParameterAnalysis}b respectively. The best performers are summarised in Fig.~\ref{fig:ParameterAnalysis}d. This action significantly improved performance on the textural FMNIST dataset, boosting accuracy from 82.0\% to 87.1\%. In stark contrast, it provided no meaningful benefit for the structural MNIST dataset (93.7\% vs. 93.3\%). This suggests that while a richer random projection is crucial for textural data, the fibre's intrinsic modal interactions were already sufficient for the simpler structural features. We then explored the reservoir's passive structure by testing fibres of varying core size (25~\textmu m, 50~\textmu m, and 62.5~\textmu m) in Fig.~\ref{fig:ParameterAnalysis}a and Fig.~\ref{fig:ParameterAnalysis}e, and length (1~m, 2~m, and 3~m) in Fig.~\ref{fig:ParameterAnalysis}b and Fig.~\ref{fig:ParameterAnalysis}f. The largest 62.5~\textmu m fibre core, which supports the greatest number of modes, yielded the highest performance for both datasets. An optimal interaction length of 1~m was identified, beyond which performance degraded for the MNIST dataset as the length increased. While the FMNIST dataset appeared to have the same optimal length, there did not seem to be the same level of degradation as the length increased. 

The most dramatic evidence of task-dependent design comes from implementing a feedback loop within the reservoir \cite{tegin2021scalable}. We implemented the feedback loop using a 2~m, 50:50, 50~$\mu$m  fibre splitter to recirculate a portion of the output signal back into the reservoir (Fig.~\ref{fig:ParameterAnalysis}c,g). This loop creates a form of \textbf{recurrent linear scattering}, a concept recently shown to be powerful for inducing complex nonlinear mappings \cite{xia2024}. In our implementation, we enrich this process using an off-the-shelf variable fibre attenuator, which operates by longitudinally separating the fibre cores with a precision screw. This creates a free-space gap where the diverging cone of light from the first fibre is clipped by the acceptance cone of the second. This mechanism preferentially attenuates higher-order modes, acting as a complex spatial filter and introducing a structural transformation within the loop. This physical enrichment explains why the system performs best with moderate (6~dB) attenuation; with no attenuation (0~dB) this filtering is absent, and at high attenuation, too much of the core signal is lost. 
\begin{figure}[tb]
    \centering
    \includegraphics[width=\columnwidth]{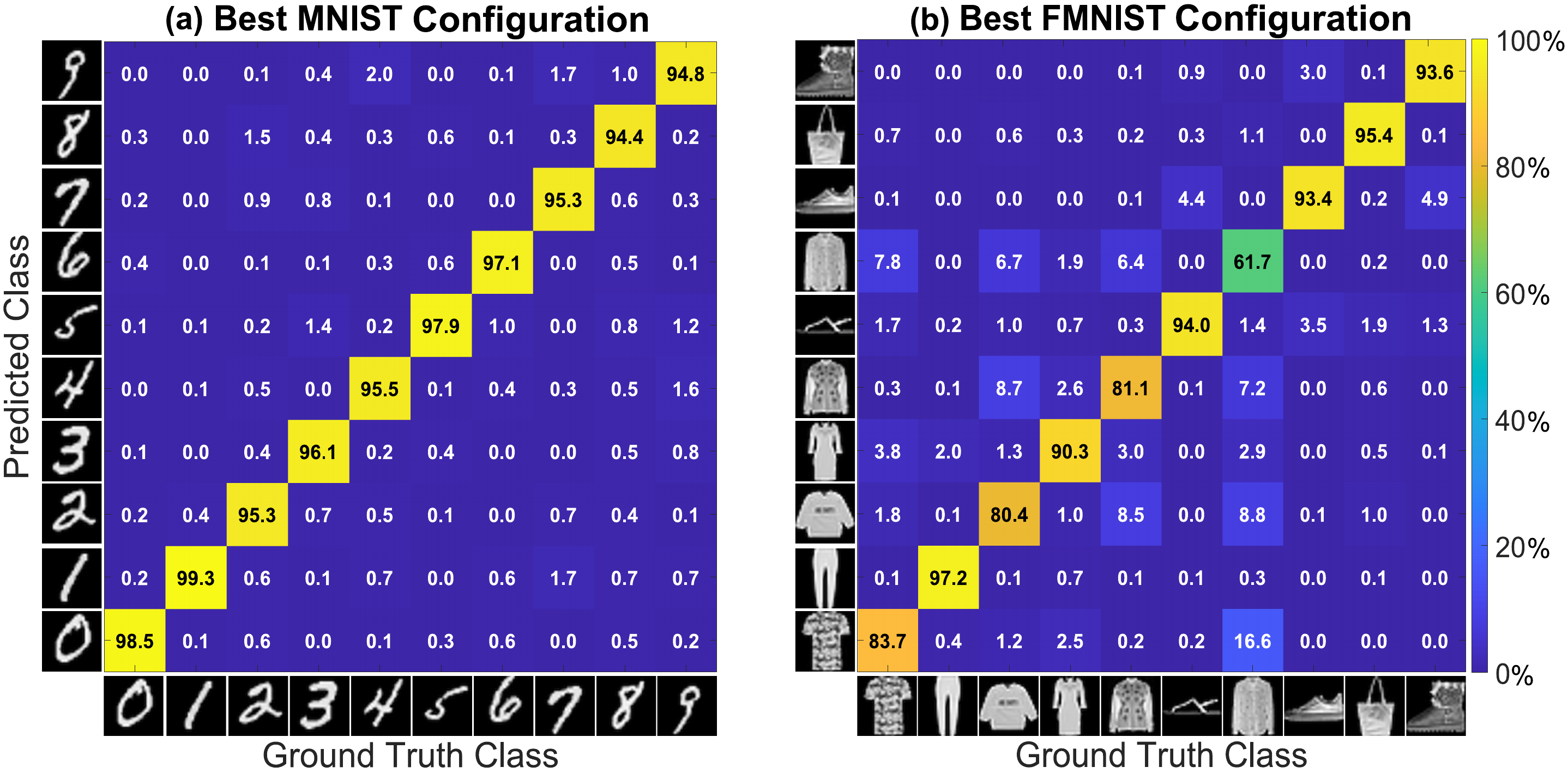} 
    \caption{\textbf{Crosstalk matrices for the best performing MNIST and FMNIST configurations.} The best performing MNIST reservoir (a) is comprised of a 2~m long, 50~$\mu$m, 50:50 fibre splitter with 6~dB of attenuation in the feedback loop (96.5~\%). The best performing FMNIST reservoir (b) is built using 1m of OM1 fibre that has increased mode mixing by submerging the fibre in weighted pellets and saturation at the camera (87.1~\%).}
    \label{fig:MFCrosstalk}
\end{figure}
\begin{figure*}[htb]
    \centering
    \includegraphics[width=\textwidth]{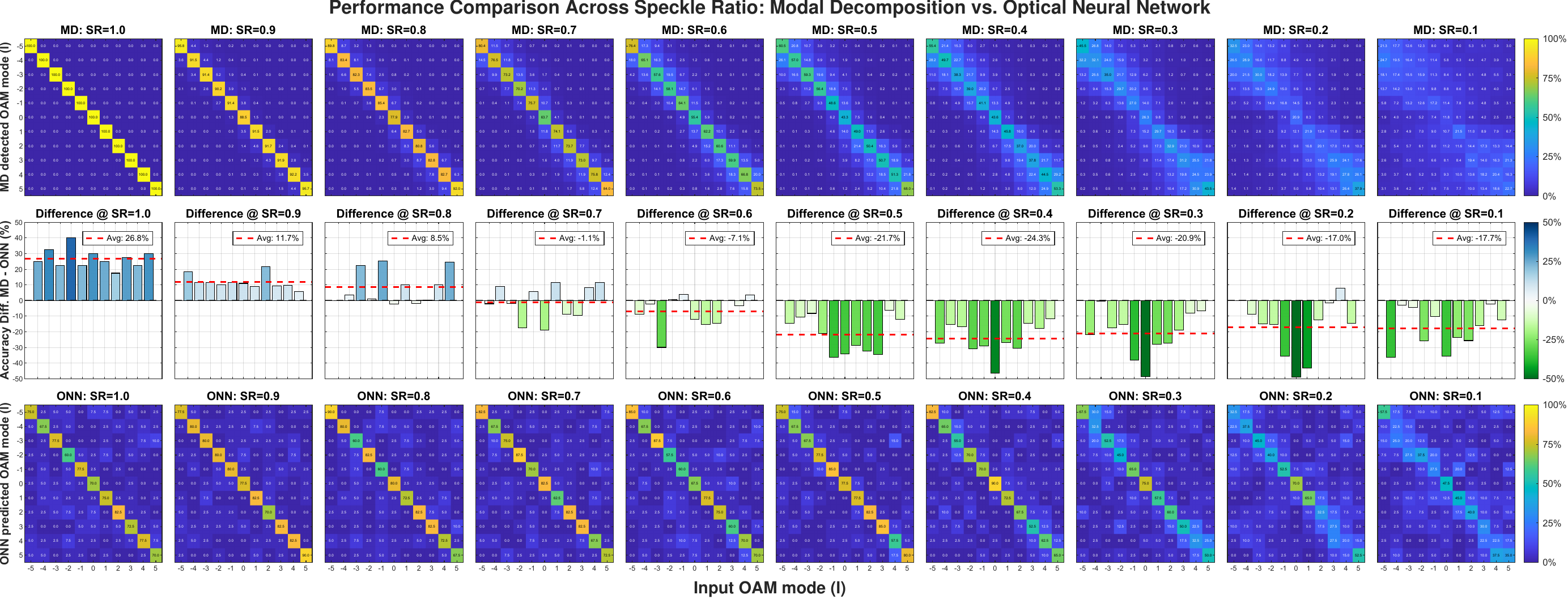} 
    \caption{\textbf{Robustness of the task-dependent optical reservoir in a turbulent environment.} The classification performance of our task-optimised optical neural network (ONN, bottom row) is compared against a conventional, power-normalised modal decomposition (MD) (top row) across a range of simulated turbulence strengths (Strehl ratio, SR). While the MD is superior in near-ideal conditions (SR $>$ 0.8), its performance collapses as turbulence increases. In contrast, the ONN maintains high fidelity. The per-mode accuracy advantage (ONN - MD), plotted in the middle row, quantifies this robustness, showing the ONN outperforms MD on average by  20.32~$\pm$~3.00~\% in the challenging regimes where classification is most needed.}
    \label{fig:OAM_Results}
\end{figure*}

This feedback provided the single largest performance increase for the structural MNIST dataset, elevating accuracy to a system-best of 96.5\%. This result was achieved using the fibre splitter feedback reservoir configuration with 6dB of attenuation. The crosstalk matrix for this configuration is shown in Fig.~\ref{fig:MFCrosstalk}a. Conversely, this same configuration was actively detrimental to the textural FMNIST dataset, with the highest classification accuracy plummeting from 87.1\% to 81.1\%, possibly because the recurrent signal transformations, while beneficial for structural data, introduced an interference or masking effect that obscured the subtle textural information for FMNIST, leading to performance degradation. Finally, we controlled the readout nonlinearity by adjusting the camera's exposure time to deliberately induce pixel saturation. The best saturation results are summarised in Fig.~\ref{fig:ParameterAnalysis}h. This hard-clipping activation was beneficial for FMNIST but offered no advantage to the peak MNIST performance. The best performing FMNIST reservoir was comprised of a 1~m OM1 fibre that had been submerged in weighted pellets to increase mode mixing and had saturated pixels. The crosstalk matrix of this configuration is shown in Fig.~\ref{fig:MFCrosstalk}b. 

These results lead to a crucial design principle: the optimal reservoir is not universal. For data defined by structure, a dynamic reservoir with recurrent features is superior. For data defined by complex textures, a spatially diverse reservoir with high mode mixing and readout nonlinearity is required. Armed with this principle, we can now select a hardware configuration specifically tailored to tackle our primary challenge of classifying phase-encoded OAM modes distorted by turbulence.

\paragraph{Robust OAM Classification in a Turbulent Environment}

\begin{figure}[ht]
    \centering
    \includegraphics[width=\columnwidth]{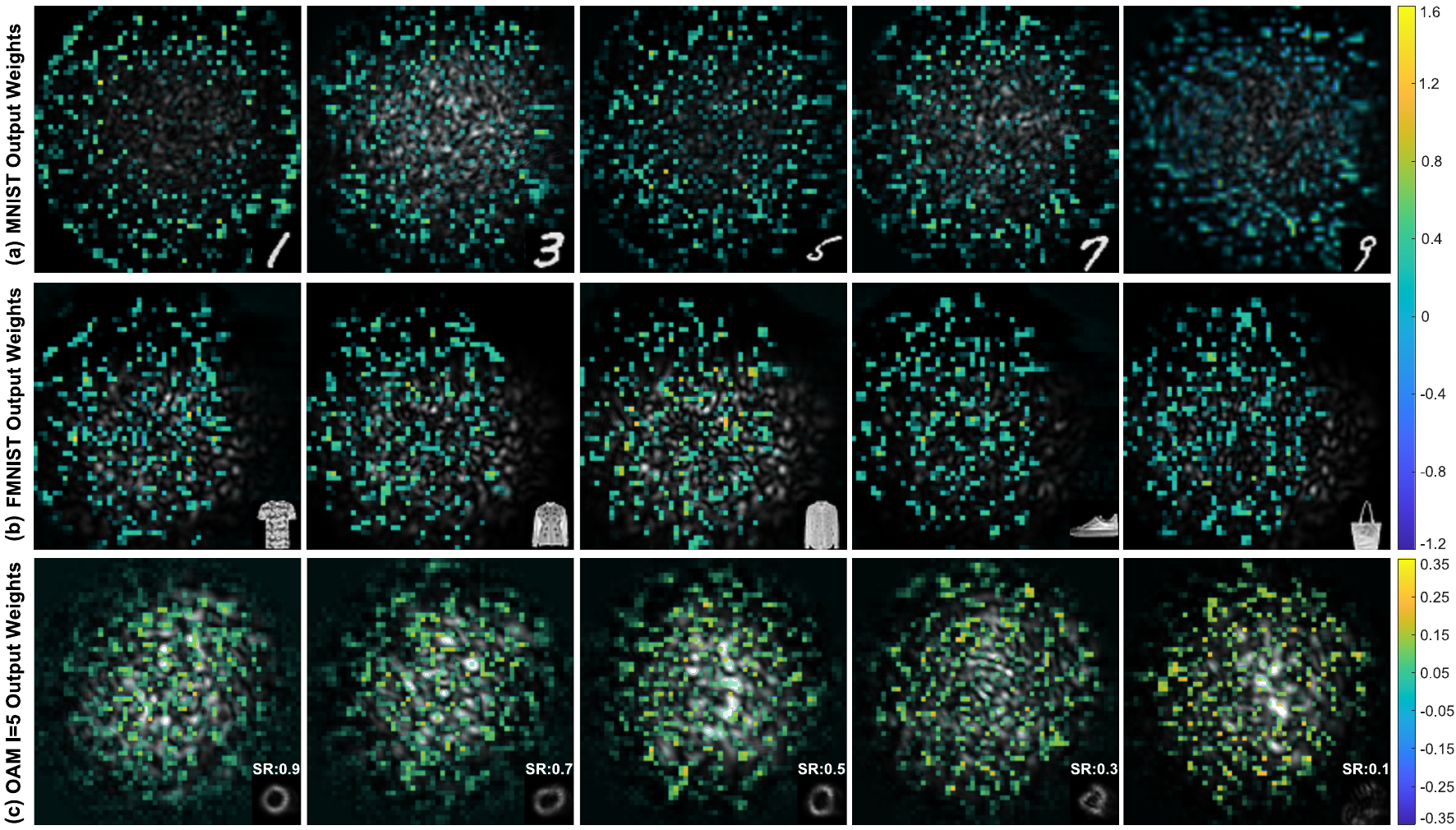} 
    \caption{\textbf{Task Dependent trained output weight distribution .} The MNIST output weights (a) are focused around the perimeter of the speckle pattern in the search of edge based features. Both the FMNIST (b) and OAM (c) output weights are distributed throughout the speckle pattern, highlighting distinctive textural and phase features.}
    \label{fig:OutputWeights}
\end{figure}

The complex, fine-grained phase and amplitude scrambling induced by turbulence is physically analogous to the textural features of the FMNIST dataset. We therefore hypothesised that the reservoir configuration optimised for that task (one with high induced mode mixing and readout nonlinearity) would be best suited for this application.

We evaluated the optimised reservoir by tasking it with the classification of 11 Laguerre-Gauss modes ($\ell \in [-5, 5]$, $p=0$). Experimentally, these modes were propagated through Kolmogorov phase screens of varying strength, ranging from a Strehl Ratio (SR) of 1.0 (no turbulence) down to 0.1. These phase screens were implemented using an SLM.  The classification was then performed on the resulting far-field intensity patterns.

To benchmark the performance of our physical system, a simulated numerical modal decomposition classifier is implemented to identify the modes affected by the same turbulence screens, also in the far field \cite{pinnell2020modal}. To ensure a fair comparison, the power measured across the 11 basis modes in this simulation was normalised to form a conditional probability distribution. This normalisation represents the optimal performance achievable by a conventional mode sorter that perfectly measures the power in each basis mode, without accounting for experimental non-idealities such as detector noise or imperfect mode projection, thereby establishing a rigorous upper bound for conventional methods under the given turbulence conditions: an ideal ``mode sorter'' \cite{Berkhout2010a,forbes2016,Ruffato2018}. 
The classification results for the simulated ideal mode sorter and the experimental optical neural network are presented in Fig.~\ref{fig:OAM_Results}a and Fig.~\ref{fig:OAM_Results}c respectively. 

The data reveals a stark trade-off between the two methods. As expected, in low-turbulence conditions ($\text{SR} > 0.8$), the conventional modal decomposition excels, offering superior classification accuracy by matching the near-perfect input modes to their ideal basis functions. However, its performance degrades sharply as turbulence increases. In contrast, our optical neural network exhibits a far greater resilience. At a Strehl ratio of 0.7, the performance of the two systems becomes comparable. For all turbulence strengths greater than this, our physical reservoir significantly outperforms the conventional method. The advantage is most pronounced in high-turbulence regimes where robust classification is most critical. Our system delivers an accuracy advantage of 22.3~$\pm$~1.78~\% for Strehl ratios between 0.5 and 0.3. When high turbulence (up to a $\text{SR} = 0.1$) is taken into account, the reservoir maintains an average 20.32~$\pm$~3.00~\% performance advantage as seen in Fig.~\ref{fig:OAM_Results}b.

These distinct output weighting patterns provide insight into the specific spatial features the optical reservoir leverages for different data structures as seen in Fig~\ref{fig:OutputWeights}. For structural data, such as MNIST, the system predominantly extracts information associated with higher spatial frequencies, which are often concentrated at the perimeter of the speckle patterns and are indicative of prominent edges. In contrast, for the more intricate textural features of FMNIST and the complex, distributed phase distortions characteristic of OAM modes in turbulence, critical information is distributed across a broader spectrum of spatial frequencies. This necessitates the recruitment of a spatially diverse ensemble of features, encompassing both low and high-order modes, to ensure robust classification.

This result demonstrates the practical power of our approach. A conventional modal decomposition relies on a rigid, one-to-one mapping of an input mode to a single basis function, a method that fails catastrophically when the input structure is corrupted. Our physical reservoir, however, is a true machine learning system: it learns to identify a distributed set of robust features within the complex speckle fingerprint that survive turbulent distortion. It is precisely this fundamentally different, data-driven approach that enables our simple, task-designed system to succeed where conventional analytical methods fail.

\section{Discussion}
\begin{table*}[tbp!]
    \centering
    \caption{Comparison of physically implemented optical neural networks, sorted by peak accuracy. The table highlights experimental configurations, the source of system nonlinearity, datasets, computational elements, and classification accuracy. All the optical neural networks had square law detection nonlinearities at the detection interface.}
    \label{tab:literature}
    \footnotesize
    \begin{tabularx}{\textwidth}{| >{\raggedright\arraybackslash\hsize=1.6\hsize}X |
    >{\raggedright\arraybackslash\hsize=0.5\hsize}X |>{\raggedright\arraybackslash\hsize=1.2\hsize}X | >{\raggedright\arraybackslash\hsize=1.2\hsize}X | >{\raggedright\arraybackslash\hsize=0.55\hsize}X | >{\raggedright\arraybackslash\hsize=0.95\hsize}X |}
        \hline
        \textbf{Implementation} & \textbf{Physical Layers} & \textbf{System Nonlinearity} & \textbf{Benchmark} & \textbf{Elements} & \textbf{Testing Accuracy} \\
        \hline
        \multicolumn{6}{|l|}{\hspace{5pt}\rule{0pt}{2.5ex}\rule[-1.2ex]{0pt}{0pt}\textbf{Diffractive Deep Neural Networks}} \\
        \hline
        3D-printed D\textsuperscript{2}NN \cite{Mengu2020} 
        & \makecell[t]{5}
        & Diffraction (structural)
        & MNIST \newline FMNIST 
        & 800\,000 \newline 800\,000 
        & 98.71\% \newline 90.04\% \\
        \hline
        SLM-based D\textsuperscript{2}NN \cite{dong2023optimized} 
        & \makecell[t]{3}
        & Electronic physical ReLU 
        & MNIST \newline FMNIST 
        & 1\,200 \newline 1\,200 
        & 97.97\% \newline 87.85\% \\
        \hline
        Photolithography D\textsuperscript{2}NN \cite{Duan2023} 
        & \makecell[t]{5}
        & Diffraction (structural)
        & MNIST ($\Phi$/$\Delta$) \newline FMNIST ($\Phi$/$\Delta$) 
        & 800\,000 \newline 800\,000 
        & 97.60\% / 97.40\% \newline 88.90\% / 88.50\% \\
        \hline
        3D-printed D\textsuperscript{2}NN \cite{lin2018all} 
        & \makecell[t]{7} 
        & Diffraction (structural)
        & MNIST 
        & 280\,000 
        & 93.39\% \\
        \hline
        Microlens-array D\textsuperscript{2}NN \cite{wang2023image} 
        & \makecell[t]{5} 
        & Saturation using an image intensifier tube
        & Organelle classification \newline QuickDraw \newline 3D scene recognition 
        & 1\,640 \newline 1\,640 \newline 1\,638 
        & 93.0\% \newline 79.0\% \newline 72.3\% \\
        \hline
        SLM-based D\textsuperscript{2}NN \cite{Zhou2020} 
        & \makecell[t]{10}
        & Ferroelectric thin films 
        & MNIST \newline MNIST back-prop. 
        & 225\,000 \newline 225\,000 
        & 92.19\% \newline 91.96\% \\
        \hline
        \multicolumn{6}{|l|}{\hspace{5pt}\rule{0pt}{2.5ex}\rule[-1.2ex]{0pt}{0pt}\textbf{Fibre-Reservoir Computing}} \\
        \hline
        CW driven MMF \cite{momeni2023backpropagation} 
        &  \makecell[t]{3 \\ 2 \\ 6} 
        & SLM phase encoding 
        & Unsupervised MNIST \newline Supervised MNIST \newline Supervised FMNIST 
        & 2\,028 \newline 1\,352 \newline 4\,056 
        & 96.51\% \newline 96.36\% \newline 87.79\% \\
        \hline
        \textbf{This Work} 
        & \makecell[t]{2}
        & Camera saturation \& physical feedback (structural)
        & MNIST \newline FMNIST \newline OAM ($\text{SR}=0.6$) \newline OAM ($\text{SR}=0.1$)
        & 72\,900 \newline 72\,900 \newline 1\,296 \newline 900
        & 96.50\% \newline 87.08\% \newline 70.68\% \newline 35.00\% \\
        \hline
        Pulsed-laser driven MMF \cite{oguz2023forward} 
        & \makecell[t]{1}
        & In-fibre Kerr effect \& digital ReLU 
        & MNIST 
        & 24\,638 
        & 94.4\% \\
        \hline
          \multicolumn{6}{|l|}{\hspace{5pt}\rule{0pt}{2.5ex}\rule[-1.2ex]{0pt}{0pt}\textbf{Photonic Chips}}\\
           \hline
        MZI circuit  \cite{zhang2024improved} 
        &  \makecell[t]{2}
        & Digital Softmax
        & MNIST 
        & 512 
        & 97.20\% \\
           \hline
        BPNC \cite{feng2022compact} 
        &  \makecell[t]{4}
        & Digital ReLu6
        & MNIST 
        & 698 
        & 94.16\% \\
        \hline
        Phase-change metasurface \cite{wu2021programmable} 
        &  \makecell[t]{3}
        & Physical ReLu
        & MNIST 
        & 1\,462 
        & 91.00\% \\
        \hline
    \end{tabularx}
\end{table*}

Current efforts in optical neural networks often focus on sophisticated algorithmic training or intricate component engineering, frequently treating the physical optical system as a fixed black box. This work, however, advocates for a significant advancement in the design philosophy for optical machine learning systems, demonstrating that the hardware itself can be intelligently configured. Diverging from approaches that necessitate complex fabrication or high-power pulsed lasers, we experimentally show that a simple, continuous-wave driven multimode fibre can be physically tailored to achieve highly competitive classification performance for specific tasks.

Our results suggest that a more integrated, task-specific approach is better suited for physical systems. By understanding and harnessing the intrinsic dynamics of the fibre, by adjusting the degree of mode mixing, the effect of feedback, and the nonlinearity at readout, we can co-design the physical system to be maximally efficient for a specific problem domain. This approach not only significantly reduces hardware complexity and power requirements, but also offers a path towards systems that are inherently robust to the noise and imperfections of the physical world.

To contextualise our findings, we compare our performance with other leading physical optical neural networks in Table~\ref{tab:literature}. The data shows that engineered Diffractive Deep Neural Networks, which require complex fabrication, achieve the highest benchmark accuracies. Our peak MNIST accuracy of 96.5\% is directly comparable to the highest-performing continuous-wave-driven reservoir reported \cite{momeni2023backpropagation}. However, these two systems represent fundamentally different design philosophies. That work treats the physical system as a fixed black box and introduces a novel, universal training algorithm to interface with it. In contrast, our work demonstrates a methodology for designing the physical hardware itself. It is crucial to distinguish this approach from the microscopic, prescriptive design of a D\textsuperscript{2}NN, where the goal is to engineer a specific mathematical transformation by calculating the phase value of millions of individual elements. Our methodology is macroscopic and parametric: we "freely" exploit the immense intrinsic complexity of the fibre and merely tune its bulk statistical properties by adjusting a handful of physical parameters which is a fundamentally simpler and more robust design approach. The key performance gain on our MNIST task did not come from an algorithmic innovation, but from a physical one: the introduction of the feedback loop. This demonstrates that by tailoring the physical reservoir's properties to the specific features of the data, state-of-the-art performance can be achieved with a computationally efficient, one-shot analytical training step. This suggests a powerful future direction where these philosophies are merged; it is plausible that applying advanced training algorithms to our physically optimised reservoirs could yield systems superior to either approach alone, reinforcing the value of task-specific hardware design as a foundational step for all physical machine learning systems.

The optical neural network's robust OAM classification performance, achieving a 20.32~$\pm$~3.00~\% improvement in accuracy over modal decomposition below a SR of 0.5, clearly demonstrates the effectiveness of even a straightforward continuous-wave fibre reservoir. This advantage is particularly pronounced in scenarios where conventional modal decomposition, while a foundational detection method in pristine conditions, fails catastrophically due to severe wavefront distortion. Another approach is to implement photonic chips, which are now also being implemented for modal decomposition \cite{sharma2025universal}. These offer the advantage of being a compact and consumer-friendly out-of-the-box solution, but the wavelength dependence and implementation of integrated photonics increase both the cost and manufacturing complexity.  However, these solutions tend to implement digital nonlinearities, increasing the power requirements and introduce a computational bottleneck into the system.

An optical approach directly circumvents the immense power and latency burdens associated with digital machine learning techniques, such as convolutional neural networks, which render them impractical for real-time FSO applications. This outcome is especially notable as our system relies solely on the intrinsic quadratic nonlinearity of square-law detection and camera saturation, which can also be achieved with high-speed photodiodes \cite{dubey2022activation}. Building on the principle of recurrent linear scattering demonstrated by our feedback loop, a subsequent step would involve replacing the simple variable attenuator with an SLM. This would enable precise, programmable control over the recurrent transformations and facilitate the introduction of structural nonlinearities within the reservoir \cite{Yildirim2024}.

This research opens several compelling avenues for future work. The simplicity of our approach, which relies on only a final linear classifier, offers a clear path towards \textit{all-optical inference}. The high-dimensional speckle patterns, already containing rich information about the input state, are directly transformed by the learned output weights. These $\mathbf{W}_{\text{out}}$ can be understood as physical matched filters which, when applied optically, could yield direct likelihoods of the input modes at the speed of light. This capability not only enables classification without electronic bottlenecks but also inherently provides the "soft information" required for state-of-the-art soft-decision forward error correction, transforming the optical front-end into an intelligent information processor. Finally, by increasing the system's interaction (input/output) speed to be faster than the loop delay, the reservoir would become a true physical Echo State Network, capable of processing temporal data and tackling time-series prediction tasks. Ultimately, our findings suggest that by harnessing and intelligently configuring the rich, natural dynamics of physical systems, we can build simpler, more efficient, and more robust computational machines.

\section{Methods}

\paragraph{Theoretical Framework}
The computational model for our physical reservoir can be understood through its analogy to digital neural networks. A reservoir without feedback is analogous to an Extreme Learning Machine (ELM), a single-layer feedforward network where only the output weights are trained. For a single input sample $\mathbf{x} \in \mathbb{R}^{1 \times N_{\text{features}}}$, where $N_{\text{features}}$ is the number of input features, the hidden state vector $\mathbf{h} \in \mathbb{R}^{1 \times N_{\text{neurons}}}$ is given by:
\begin{equation}\label{eq:hELM}
    \mathbf{h} = f(\mathbf{x} \mathbf{W}_{\text{in}} + \mathbf{b}),
\end{equation}
where $N_{\text{neurons}}$ is the number of neurons in the reservoir, $\mathbf{W}_{\text{in}} \in \mathbb{R}^{N_{\text{features}} \times N_{\text{neurons}}}$ is the fixed, randomised input weight matrix, $\mathbf{b}$ is a bias vector, and $f$ is a nonlinear activation function. For a batch of $N_{\text{samples}}$ inputs, the hidden states are stacked to form the matrix $\mathbf{H} \in \mathbb{R}^{N_{\text{samples}} \times N_{\text{neurons}}}$ \cite{huang2006extreme}.

In our physical system, the abstract numerical input features $\mathbf{x}$ (as defined in Equation \eqref{eq:hELM}) are first encoded onto a physical optical field by the spatial light modulator. This optical input field, denoted as $\mathbf{E}_{\text{in}}$, then interacts with the multimode fibre. The fibre's complex transmission properties collectively serve the role of the fixed weight matrix $\mathbf{W}_{\text{in}}$. The nonlinearity $f$ is not programmed but arises physically from the square-law intensity detection and eventual saturation of the camera. The captured physical state of our reservoir is thus given by:
\begin{equation}\label{eq:resPhysical}
    \mathbf{h} = \min(|\mathbf{E}_{\text{out}}|^2, L_{\text{sat}}), \quad \text{where} \quad \mathbf{E}_{\text{out}} = \mathbf{E}_{\text{in}} \mathbf{T},
\end{equation}
where $\mathbf{E}_{\text{out}}$ is the complex optical field at the camera plane and $L_{\text{sat}}$ is the camera's saturation level.

When a feedback loop is introduced, the system can be modelled as an Echo State Network (ESN), which includes memory. In the general case, the hidden state $\mathbf{h}_n$ at time step $n$ depends on the previous state $\mathbf{h}_{n-1}$ and the current input $\mathbf{x}_n$:
\begin{equation}
    \label{eq:hESN}
    \mathbf{h}_n = (1-\gamma)\mathbf{h}_{n-1} + \gamma f(\mathbf{x}_n \mathbf{W}_{\text{in}} + \mathbf{h}_{n-1} \mathbf{W}_{\text{loop}}),
\end{equation}
where $\mathbf{W}_{\text{loop}}$ is the recurrent weight matrix and $\gamma$ is a leak rate \cite{lukosevicius2012practical}. As our system operates in a "slow" regime, where the interaction time is longer than the loop delay, it does not exhibit memory but benefits from enhanced spatial complexity. 

Crucially, the variable attenuator in the loop is not a simple scalar loss. It is a complex physical element that induces mode-dependent losses (spatial filtering) and random polarisation transformations due to misalignment. This is better described by a matrix operator, $\mathbf{A}_{l}$, which encapsulates these effects. Considering the feedforward path transmission matrix as $\mathbf{T}_{\text{FF}} = \mathbf{T}_{\text{A1}} \cdot \mathbf{T}_{\text{Splitter}} \cdot \mathbf{Q}_{\text{1}} \cdot \mathbf{T}_{\text{B1}}$, and the loop path transmission matrix as $\mathbf{T}_{\text{Loop}} = \mathbf{T}_{\text{Splitter}} \cdot \mathbf{Q}_{\text{2}} \cdot \mathbf{T}_{\text{B2}} \cdot \mathbf{A}_{l} \cdot \mathbf{T}_{\text{A2}}$, the effective transmission matrix of the reservoir, taking into account the feedback, can then be described as:
\begin{equation}\label{eq:TMeff}
    \mathbf{T}_{\text{Eff}} = \mathbf{T}_{\text{FF}} \cdot (\mathbf{I} - \mathbf{T}_{\text{Loop}})^{-1},
\end{equation}
where $\mathbf{I}$ is the identity matrix of appropriate dimensions. The $\mathbf{T}$ terms represent the transmission matrices of the respective fibre and splitter components, $\mathbf{Q}$ denote the splitting ratios, and $\mathbf{A}_{l}$ is the transmission matrix of the loop operator, which accounts for both attenuation and mode filtering. 


\paragraph{Reservoir Parameter Characterisation}
To establish our design principles, we systematically varied six key physical parameters. We tested three \textbf{fibre lengths} of OM1 (1~m, 2~m, and 3~m) and three \textbf{fibre core types} (25~\textmu m step-index, 50~\textmu m step-index, and 62.5~\textmu m graded-index OM1). We controlled \textbf{mode mixing} by applying weighted pellets to the fibre coil to induce micro-bends. A \textbf{feedback loop} was implemented using a 2~m long, 50~\textmu m core, 50:50 fibre splitter connected to an off-the-shelf \textbf{variable fibre attenuator}, which operates by longitudinally separating the fibre cores with a precision screw, providing the complex mode-filtering behaviour. Finally, we studied \textbf{readout nonlinearity} by adjusting the camera's exposure time to achieve either a low-saturation or high-saturation regime. These configurations were benchmarked using the standard MNIST and FMNIST datasets with their conventional 60,000 training and 10,000 testing item splits.

\paragraph{Training and Evaluation}
The speckle patterns captured by the camera form the rows of the hidden state matrix $\mathbf{H}$. The output weights, $\mathbf{W}_{\text{out}} \in \mathbb{R}^{N_{\text{neurons}} \times N_{\text{classes}}}$, are then computed in a single analytical step using ridge regression for its numerical stability \cite{lukosevicius2012practical}. This method was chosen after preliminary tests showed it to be more robust for this application than other linear models, such as Lasso regression. Given a one-hot encoded target matrix $\mathbf{Y} \in \mathbb{R}^{N_{\text{samples}} \times N_{\text{classes}}}$, the output weights are calculated as:
\begin{equation}\label{eq:RidgeRegression}
    \mathbf{W}_{\text{out}} = (\mathbf{H}^{\top} \mathbf{H} + \rho \mathbf{I})^{-1} \mathbf{H}^{\top} \mathbf{Y},
\end{equation}
where $\rho$ is the Tikhonov regularisation coefficient (set to 1) and $\mathbf{I}$ is the identity matrix. Predictions for unseen test data, $\mathbf{H}_{\text{test}}$, are then given by $\mathbf{Y}_{\text{pred}} = \mathbf{H}_{\text{test}} \mathbf{W}_{\text{out}}$. The number of computational elements, $N_{\text{neurons}}$, corresponds to the number of camera pixels used. Performance scaling was tested by binning adjacent camera pixels before training.

\paragraph{OAM Classification Benchmark}
To evaluate the system's real-world utility, the optimised reservoir was tested on the classification of 11 Laguerre-Gaus  modes (radial index p=0, azimuthal index $\ell \in [-5, 5]$) \cite{pinnell2020modal}. This dataset was generated by distorting each mode with 200 unique Kolmogorov turbulence screens at 10 distinct strengths, corresponding to Strehl ratios of 1.0 (no turbulence) down to 0.1: a total of 22\,000 measurements. Both the modes and the turbulence phase screens were encoded onto the SLM. The input is transformed into the far field by the Fourier transformation used to focus the input beam into the fibre.

The performance of the optical neural network was benchmarked against an idealised numerical modal decomposition. To ensure a fair comparison, the power measured across the 11 basis modes in the numerical model was normalised to form a conditional probability distribution, effectively transforming the numerical crosstalk matrix into a confusion matrix. The same experimentally implemented phase screens are used in the simulation and are propagated to the far field using Fourier transformations. 

\section{Data Availability}
The data that support the findings of this study are available from the corresponding author upon reasonable request.

\section{Author Contributions}
C.R.R.: Conceptualization, Methodology, Software, Validation, Formal analysis, Investigation, Writing – Original Draft. M.A.C.: Conceptualization, Supervision, Funding acquisition, Writing – Review \& Editing.

\section{Acknowledgements}
This work was supported by the National Research Foundation (NRF) of South Africa (Grant No. TTK2204011621).

\bibliographystyle{naturemag} 
\bibliography{references}    

\end{document}